\documentclass[preprint,12pt]{elsarticle}
\usepackage{subfigure}
\usepackage{amsmath}
\usepackage{amssymb}
\usepackage{url}
\biboptions{sort&compress}
\usepackage[a4paper,total={6in, 8in}]{geometry}
\usepackage{atbegshi}
\AtBeginDocument{\AtBeginShipoutNext{\AtBeginShipoutDiscard}}

\journal{Physics Letters A}

\begin{document}

\begin{frontmatter}

\title{Road to zero-field antiferromagnetic skyrmions in a frustrated AFM/FM heterostructure}
\author[1]{M. Mohylna},
\author[1]{V. Tkachenko},
\author[1]{M. \v{Z}ukovi\v{c}\corref{cor1}}
\ead{milan.zukovic@upjs.sk}
\address[1]{Department of Theoretical Physics and Astrophysics, Institute of Physics, Faculty of Science, Pavol Jozef \v{S}af\'arik University in Ko\v{s}ice, Park Angelinum 9, 041 54 Ko\v{s}ice, Slovak Republic}
\cortext[cor1]{Corresponding author}



\begin{abstract}
We demonstrate a mechanism of significant reduction, including complete elimination, of the external magnetic field required for the stabilization of a skyrmion lattice (SkX) phase in a frustrated triangular Heisenberg antiferromagnet (AFM) with the Dzyaloshinskii-Moriya interaction. It is achieved by coupling of such a AFM plane to a reference ferromagnetic (FM) layer, which generates an effective field cooperating with the external magnetic field. If the FM layer shows some axial single-ion anisotropy then the effective field can also be generated in zero external field due to a spontaneous FM long-range ordering. Then a sufficiently large interlayer coupling can fully substitute the external magnetic field and the SkX phase in the AFM layer can be stabilized even in zero external field.
\end{abstract}

\begin{keyword}
Heisenberg antiferromagnet \sep Geometrical frustration \sep Skyrmion lattice \sep Zero-field skyrmions \sep AFM/FM bilayer
\end{keyword}


\end{frontmatter}

\section{Introduction}
Magnetic skyrmions, the whirled topological spin configurations with non-trivial magnetic properties, have been actively investigated since their experimental observation in 2009 \cite{muhlbauer2009skyrmion}. Although the earliest theoretical and experimental works mostly focused on finding the skyrmion (SkX) phase and individual skyrmions in ferromagnets (FM) \cite{belavin1975metastable, bogdanov1989thermodynamically, bogdanov1994thermodynamically,muhlbauer2009skyrmion, do2009skyrmions, heinze2011spontaneous}, more recently the attention has been shifted to antiferromagnetic (AFM) materials, the advantages of which include higher robustness and elimination of undesirable effects in the skyrmion current \cite{zhang2016antiferromagnetic, barker2016static, bessarab2019stability}.

One of the typical mechanisms that can lead to the stabilisation of skyrmions of both Ne\'el and Bloch types is the Dzyaloshinskii-Moriya interaction (DMI) \cite{dzyaloshinsky1958thermodynamic, moriya1960anisotropic}, which breaks the inversion symmetry in the system. Several studies have focused on the frustrated AFM triangular Heisenberg magnet with the presence of DMI \cite{rosales2015three, osorio2017composite, mohylna2021formation, mohylna2022stability, fang2021spirals}. Rosales~\emph{et~al.} and Osorio~\emph{et~al.} by means of both Monte Carlo (MC) simulations and low-energy effective action method demonstrated, that the for a moderate strength of the DMI in such model the SkX phase is stabilised in a quite wide temperature-field range \cite{rosales2015three, osorio2017composite}. Following works investigated the evolution of the phase diagram topology of such model with varying DMI strength~\cite{mohylna2021formation} and the presence of single-ion anisotropy \cite{mohylna2022stability, fang2021spirals}. It was shown, that the increase of the DMI as well as the presence of small in-plane and out-of-plane anisotropy can to some extent widen/shrink or shift the SkX phase area both in field and temperature parameter ranges. 

Nevertheless, stabilization of the skyrmion lattice in the model requires a giant perpendicular external magnetic field, which may be difficult or impossible to achieve in the laboratory and it is certainly not practical for technological applications. For example, it is estimated that to access the SkX phase in the experimental realization of the present model in the compound Fe/MoS\textsubscript{2} the required magnetic field could be as high as 20 T~\cite{fang2021spirals}. Furthermore, the presence of more realistic features, such as strong single-ion anisotropy \cite{mohylna2022stability} and impurities~\cite{impur2022,silva2014emergence}, can lead to the further shrinking of the field range of the SkX phase stability.

An elegant approach to stabilization of the skyrmion state at much lower, including zero, external fields has been proposed by designing heterostructures composed of bilayers or multilayers~\cite{yu2018room, guang2020creating, nandy2016interlayer}. In particular, Nandy~\emph{et~al.}~\cite{nandy2016interlayer} demonstrated, that coupling the thin chiral magnet, capable of hosting the skyrmions at gigantic external magnetic field values, to a stiff FM layer with strong out-of-plane anisotropy, can lead to the stabilisation of the atomic-scale skyrmions at significantly lower fields. It was found that the effective field of the reference FM layer can even fully substitute the required magnetic field for skyrmion formation. To achieve it the reference layer should be a hard ferromagnet with a large exchange stiffness and a strong out-of-plane anisotropy, and the effective field induced by the interlayer exchange coupling should be in the range of the magnetic field required for skyrmion stabilization. The required conditions in such a heterostructure can be rather easily secured by varying the thickness of the layers and the composition at the interface. The resulting effective out-of-plane magnetic fields can be really impressive reaching above 40 T in the Mn/W\textsubscript{m}/Co\textsubscript{n}/Pt/W(001) multilayer system with 5 layers of W. This approach is rather general and can be used for any two-dimensional chiral magnet with a surface or interface induced DMI.


In the present Letter, we propose a similar mechanism that can bring the external magnetic field required for the SkX phase stabilization in the studied frustrated AFM triangular Heisenberg magnet with the presence of DMI to lower or even zero fields.

\section{Model and Method}

In the following we consider a heterostructure in the form of a bilayer consisting of the Heisenberg AFM triangular lattice with the DMI, capable of hosting the SkX phase, coupled ferromagnetically to another reference FM layer with out-of-plane anisotropy. The considered AFM/FM bilayer model is described by the Hamiltonian

\begin{equation}
\begin{split}
\mathcal{H} = & - J_{AFM} \sum_{\langle i,j \rangle}\vec{S_{i}}\cdot
\vec{S_{j}} + \sum_{\langle i,j \rangle}\vec{D_{ij}}\cdot\Big [\vec{S_{i}}\times\vec{S_{j}} \Big] \\
& - J_{FM} \sum_{\langle \alpha,\beta \rangle}\vec{S_{\alpha}}\cdot\vec{S_{\beta}} + A\sum_{\alpha} (S_{\alpha}^{z})^2 \\
& - J_{int} \sum_{\langle i,\alpha \rangle}\vec{S_{i}}\cdot\vec{S_{\alpha}} - h(\sum_i S_i^{z}+\sum_{\alpha}  S_{\alpha}^{z}),
\end{split}
\label{HamiltBI}
\end{equation}
where spins $\vec{S_i}$ and $\vec{S_j}$ belong to the AFM and $\vec{S_{\alpha}}$ and $\vec{S_{\beta}}$ to the FM layers. The first two terms in the Hamiltonian describe the AFM subsystem, with $J_{AFM} < 0 $ denoting the AFM exchange coupling constant and $\vec{D}_{ij}=D\vec{r}_{ij}/|\vec{r}_{ij}|$ being the DMI vector chosen along the radius vector between two neighbouring sites $i$ and $j$, whose strength is defined by the value of $D$. The next two terms include the FM exchange interaction with the coupling constant $J_{FM} > 0 $ and the single-ion anisotropy of the strength $A > 0$. The interlayer FM exchange interaction ($J_{int} > 0$) and the external magnetic field $h$ applied perpendicular to the lattice planes (along the $z$-direction) are expressed in the last two terms. In the following, we set $J_{AFM}=-1$, $J_{FM}=1$, and the Boltzmann constant $k_B=1$, thus fixing the energy scale.

In order to identify the region of the SkX phase stability in the AFM layer we can compute several useful quantities. As a suitable order parameter for the SkX phase we can consider the skyrmion chirality $\kappa$, the discretization of a continuum topological charge~\cite{berg1981definition}, which provides the information about the number and the nature of topological configurations present in the system. Alternatively, one can also consider the skyrmion number, providing similar information~\cite{rosales2015three,mohylna2021formation}. Abrupt changes in the order parameters at low temperatures can serve to locate the SkX phase boundaries. However, at higher temperatures their decay is rather smooth and thus the peaks in the response functions, such as the specific heat $c$, are more suitable for locating the transition points. The skyrmion chirality can be obtained as follows:

\begin{equation}
\kappa = \frac{\langle K \rangle}{N} = \frac{1}{8\pi N} \Big\langle \sum_i \Big( \kappa^{12}_{i} + \kappa^{34}_{i} \Big)\Big\rangle,
\label{chiral}
\end{equation}




where $\langle \hdots \rangle$ denotes the thermal average, $N=L^2$ is the number of spins the the layer, and $\kappa^{ab}_{i} = \vec{S_i}\cdot[\vec{S_a}\times\vec{S_b}]$ is the chirality of a triangular plaquette consisting of three neighbouring spins. Spins are taken in counter-clockwise fashion to keep the sign in accordance with the rules in Ref.~\cite{berg1981definition}. The specific heat in the AFM plane can be calculated as

\begin{equation}
c = \frac{\langle \mathcal{H}_{AFM}^2\rangle  - \langle \mathcal{H}_{AFM} \rangle^2 }{NT^2},
\label{spec_heat}
\end{equation}
where $\mathcal{H}_{AFM}$ refers to the Hamiltonian pertaining to the AFM plane.

In the FM layer, it is useful to calculate the magnetization $m_{FM}$, which is the order parameter of the FM phase, defined as
\begin{equation}
m_{FM} = \frac{\langle M_{FM} \rangle}{N} = \frac{1}{N} \Big\langle \sum_{\alpha} S_{\alpha}^z\Big\rangle.
\label{magn}
\end{equation}

The calculations are performed using the  hybrid MC (HMC) method, which combines the standard Metropolis algorithm with the over-relaxation method \cite{creutz1987overrelaxation}. The latter is a deterministic energy-preserving perturbation method, which helps decorrelate the system and leads to faster relaxation. Each layer consists of $N = L^2$ sites with the periodic boundary conditions imposed in the $x-y$ plane. We set $L = 48$\footnote{We tested several values of the lattice size and found that larger values of $L$ produced similar results.} and use up to $10^6$ MC steps (MCS) for equilibration and subsequently half of that for the calculation of the mean values. Due to the massive computational requirements of the problem, which required simulations in a relatively broad parameter space, the algorithm was parallelized on General Purpose Graphical Processing Units (GPGPU) using CUDA. 



\section{Results}
The phase diagram of a single AFM layer in the $h-T$ parameter plane for $D=0.5$ has been presented in Ref.~\cite{rosales2015three}. The AFM-SkX phase appears at low temperatures in the field range of $2 \lesssim h \lesssim 6.5$\footnote{We note that this range of the SkX phase is somewhat overestimated and later studies revealed the presence of another, so-called sublattice-uniform (SU)~\cite{osorio2017composite} or $V$-like (VL)~\cite{mohylna2022stability} phase, which appears wedged between the SkX and fully polarized FM phases.} and extends to the temperatures slightly above $T = 0.3$. This case corresponds to the situation in the AFM layer of the present bilayer model with the parameters $D = 0.5$ and $J_{int}=0$. Note that the values of the parameters $J_{FM}$ and $A$ of the FM layer are irrelevant for the AFM SkX phase, since the two layers are completely decoupled and cannot affect each other's behavior.

\begin{figure}[t!]
\centering
\subfigure{\includegraphics[scale=0.55,clip]{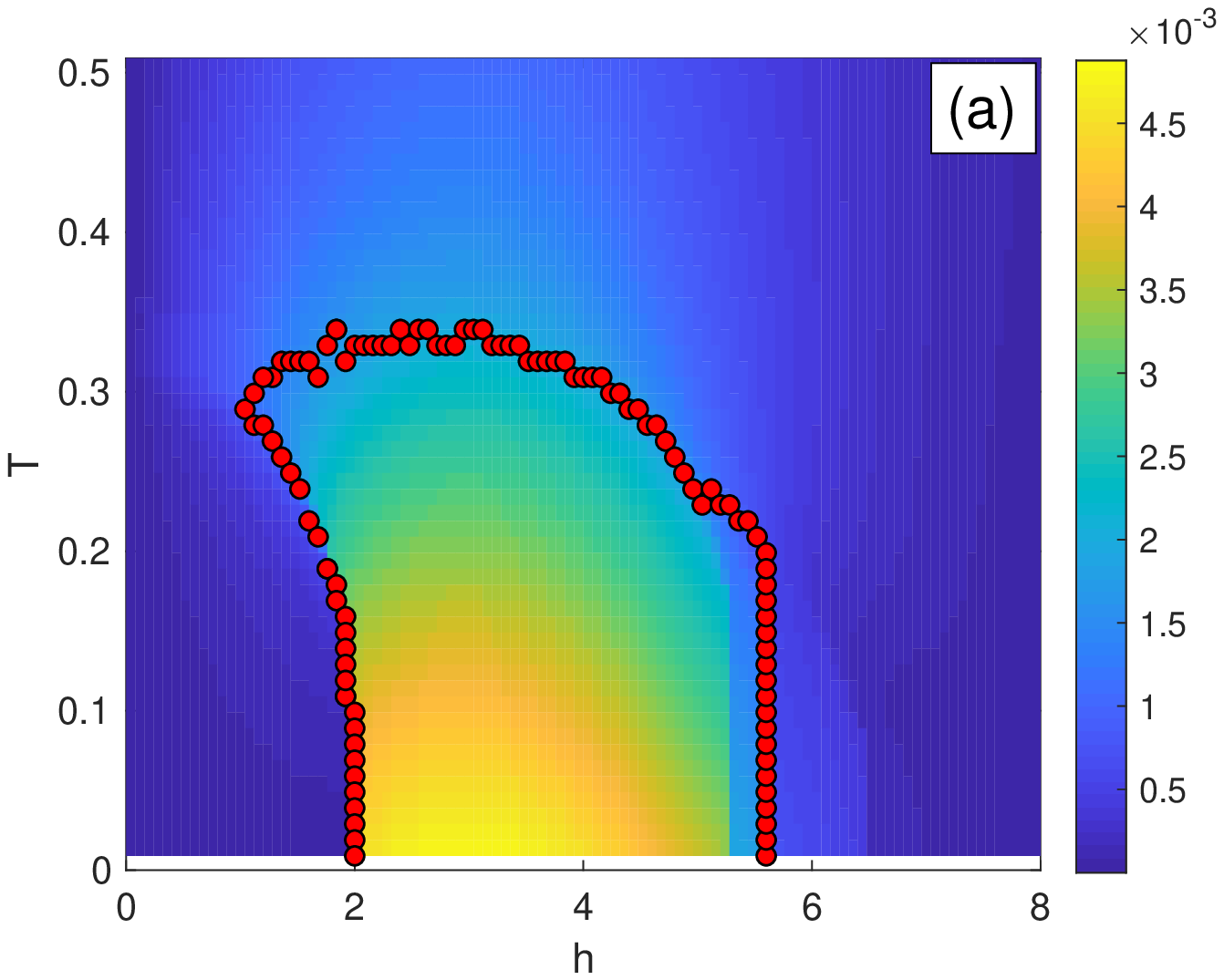}\label{fig:h-T_Jint_0}}
\subfigure{\includegraphics[scale=0.55,clip]{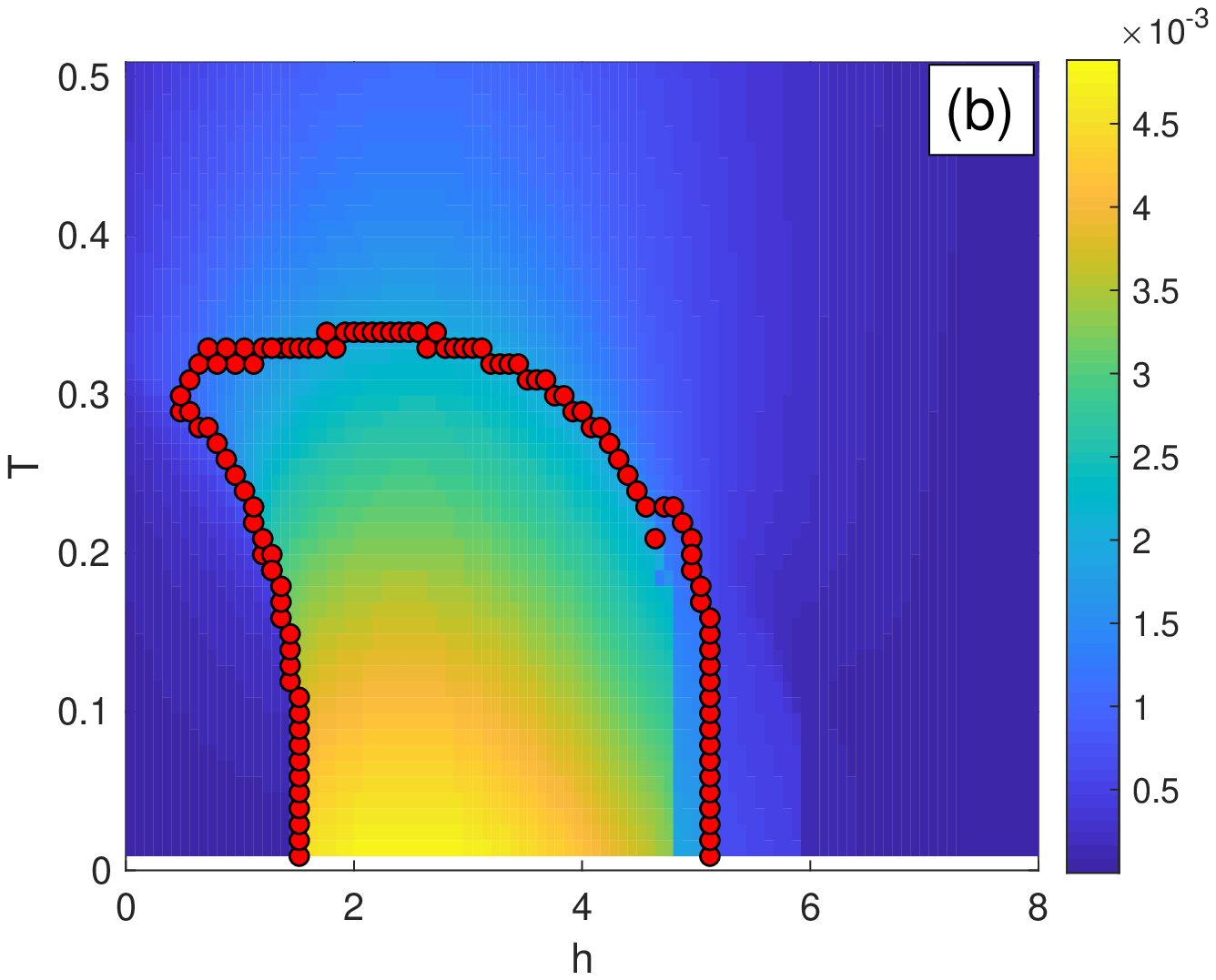}\label{fig:h-T_Jint_0_5}}
\subfigure{\includegraphics[scale=0.55,clip]{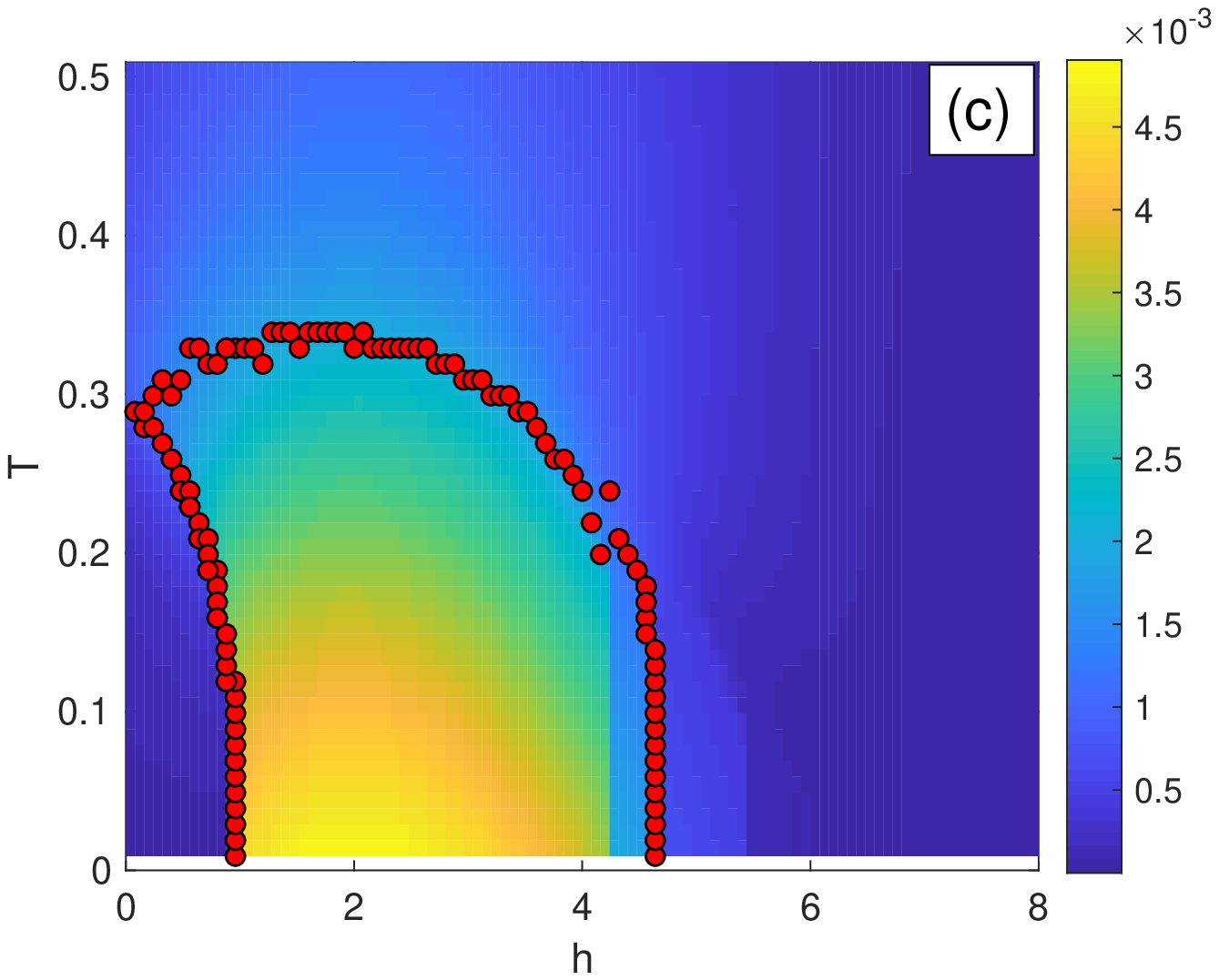}\label{fig:h-T_Jint_1_0}}
\subfigure{\includegraphics[scale=0.55,clip]{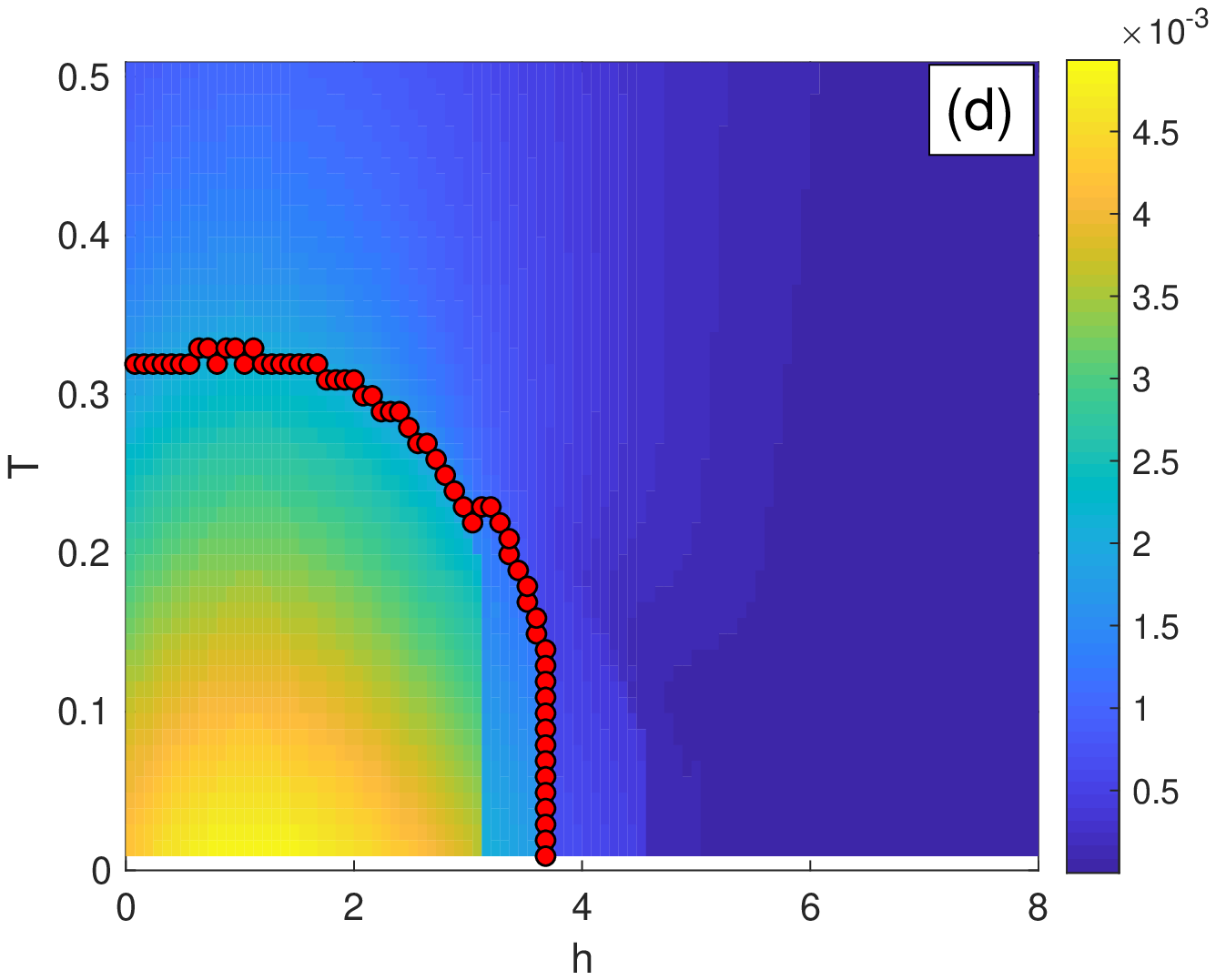}\label{fig:h-T_Jint_2_0}}
\caption{The skyrmion chirality of the AFM layer, $\kappa$, plotted in the $h-T$ parameter plane, for (a) $J_{int}=0$, (b) $0.5$, (c) $1.0$ and (d) $2.0$, with the remaining parameters fixed to $D=0.5$, $A=0.5$ and $L=48$. The red circles mark the area of the SkX phase.}
\label{fig:h-T}
\end{figure}

It is expected that when the two layers become coupled via the FM interaction $J_{int}$, with the increasing intensity of the coupling each layer will try to enforce more and more its preferred state on the other layer. Here, we are primarily interested in the effects of the FM layer imposed on the SkX phase-hosting AFM one. In the presence of the external magnetic field the decoupled FM plane will show a field-induced FM long-range ordering (LRO) even in the absence of the single-ion anisotropy. The AFM plane, when coupled to such a FM plane feels an additional effective field trying to align its spins along its preferred, i.e. the external field direction. Thus, the effective field will cooperate with the external one and its increasing intensity, mediated by $J_{int}$, can be expected to gradually alleviate demands on the latter to stabilize the SkX phase. Consequently, the window of the external fields required for the SkX appearance can be shifted to lower values. Of course, due to the presence of different competing interactions, the situation is more complex and such a scenario is not so obvious. For example, at lower fields the field-induced FM LRO in the FM layer might be destroyed by the strong coupling to the frustrated AFM layer, which prefers different kind of ordering. On the other hand, at lower fields the FM layer LRO can be reinforced by the axial anisotropy $A>0$.

In Fig.~\ref{fig:h-T} we show the evolution of the SkX phase in the $h-T$ phase diagram of the AFM layer for the increasing intensity of the interlayer coupling $J_{int}$, with the fixed $A=0.5$. The red circles, determined based on the low-temperature chirality behavior and high-temperature peaks in the response functions, mark the area of the SkX phase. As expected, for $J_{int}=0$ (Fig.~\ref{fig:h-T_Jint_0}) the phase diagram agrees well with that of the single layer~\cite{rosales2015three}. However, with the increasing $J_{int}$ the SkX phase gradually shifts to lower fields. In particular, the lower bound $h_{lb}$ of the SkX phase field interval close to $T=0$ reduces by the value approximately equal to $J_{int}$. Consequently, the SkX phase can emerge at the fields $h_{lb} \approx 1.5$ for $J_{int}=0.5$ (Fig.\ref{fig:h-T_Jint_0_5}) or $h_{lb} \approx 1$ for $J_{int}=1$ (Fig.\ref{fig:h-T_Jint_1_0}). Figure~\ref{fig:h-T_Jint_2_0} suggests that for $J_{int}=2$ the SkX phase can be stabilized even in zero external field. The upper bound $h_{ub}$ of the low-temperature SkX phase field interval also decreases by about the same value, thus preserving the total SkX phase external field interval length. On the other hand, the temperature window of the SkX phase does not seem to be affected by the value of $J_{int}$, as long as the other model parameters remain fixed.

\begin{figure}[t!]
\centering
\subfigure{\includegraphics[scale=0.58,clip]{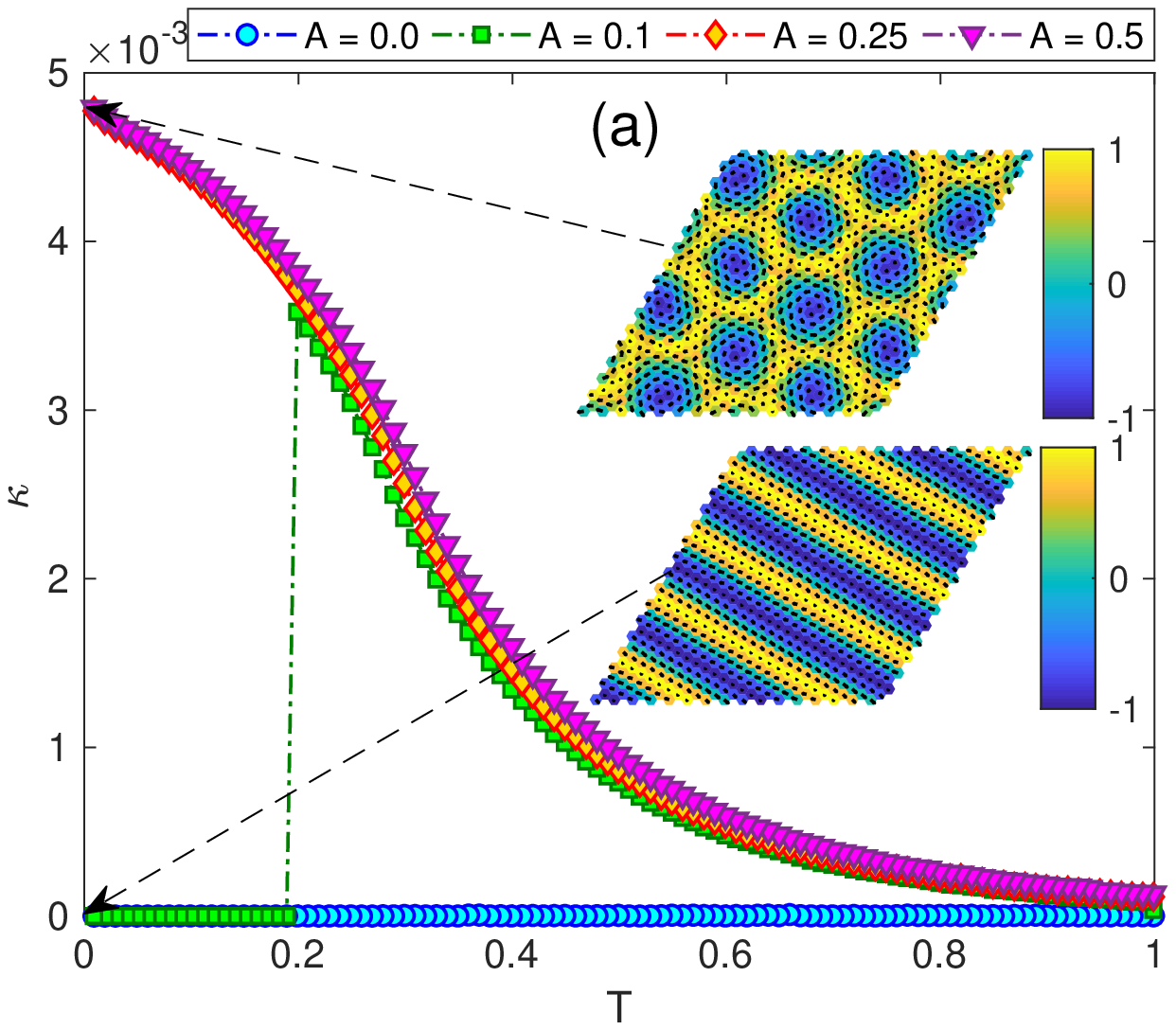}\label{fig:kappa_afm}}
\subfigure{\includegraphics[scale=0.58,clip]{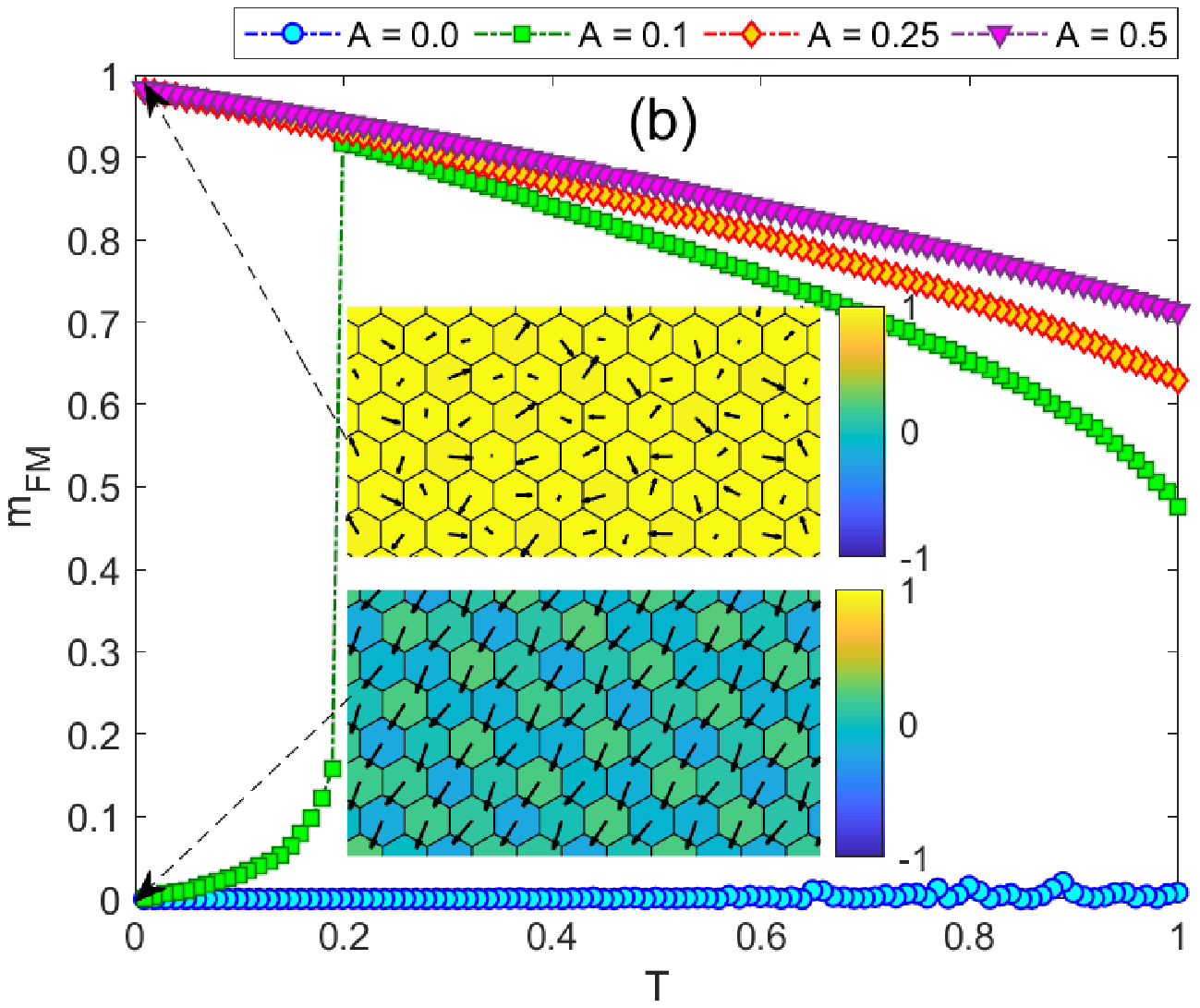}\label{fig:m_fm}}
\caption{Temperature dependencies of (a) the chirality in the AFM layer and (b) the magnetization in the FM layer, for $h=0$, $J_{int}=2$ and different values of $A$. The snapshots taken at the lowest simulated temperature demonstrate the helical state in the AFM layer imposed by the in-plane FM ordering in the FM layer for $A=0.1$ (green curves) changing to the SkX phase in the AFM layer induced by the out-of-plane FM ordering in the FM plane for $A=0.5$ (magenta curves). The arrows in the snapshots represent spin orientations in the $x-y$ plane and the colors signify the values of their $z$-component, $S^z$.}
\label{fig:zero-field}
\end{figure}

To verify the stability of the zero-field SkX phase and its dependence on the model parameters, in Fig.~\ref{fig:kappa_afm} we plot the chirality (the SkX phase order parameter in the AFM plane) and the FM layer magnetization (the FM phase order parameter in the FM plane) for different values of the single-ion anisotropy $A$ in the FM layer. As one can see, for $A=0$ both the chirality and the magnetization stay zero at all temperatures and therefore there is no SkX phase in the AFM layer that would be induced by the FM ordering in the reference FM plane. LRO in the 2D isotropic FM Heisenberg model is forbidden by the Mermin-Wagner theorem~\cite{mermin1966absence}. In the 2D anisotropic Heisenberg model with $A>0$, LRO is possible and is expected to appear at very low temperatures with the increasing $A$ gradually extend to higher temperatures~\cite{torelli2019high}. 

However, coupling of such either isotropic or anisotropic FM layer to another frustrated AFM layer makes the situation more involved in both layers. We note that critical properties of such a AFM/FM bilayer have already been studied for the special case of $A \to \infty$ (Ising model)~\cite{vzukovivc2016critical,vzukovivc2017ordering}. In the present AFM/FM bilayer system we can observe that the FM LRO starts forming in the FM layer at finite $A$, however, not at very low temperatures but only within some range of intermediate values (see the green curve in Fig.~\ref{fig:m_fm} for $A=0.1$). This FM LRO in the FM layer imposes the SkX lattice phase in the AFM plane. Nevertheless, the temperature range of the SkX phase stabilization does not coincide with the FM layer LRO phase. It seems to emerge at the low-temperature bound of the FM phase at $T \approx 0.2$ but as the temperature is increased it disappears well below the transition temperature of the FM LRO to the paramagnetic (P) phase.

For sufficiently large values of the single-ion anisotropy, such as $A=0.25$ or $0.5$ both the FM and SkX phases extend down to zero temperature. The increasing $A$ increases the FM-P transition temperature to the Ising value for $A \to \infty$~\cite{torelli2019high}, nevertheless, the SkX-P transition temperature in the AFM layer appears to be little sensitive to the value of $A$.

\section{Conclusion}
Finally, in this paper we suggested a mechanism of improving the field stability area of the SkX phase appearing in a triangular Heisenberg AFM with the DMI. We demonstrated, that FM coupling of such a AFM plane to a reference FM layer can shift the SkX phase to significantly lower fields. The shift results from the presence of an effective field generated by the FM LRO in the reference layer, which cooperates with the external magnetic field. The magnitude of the shift is proportionate to the strength of the FM interlayer exchange. If the FM layer possesses strong enough out-of-plane (axial) single-ion anisotropy then FM LRO can develop in the FM layer even in zero field. Then, if the interlayer coupling is sufficiently large the FM layer-imposed effective field can fully substitute the external magnetic field and the SkX phase can be stabilized in the AFM layer even in zero field. We believe that such a mechanism can be experimentally realized in the heterostructure involving transition metal dichalcogenides with magnetic transition metals, such as Fe/MoS\textsubscript{2}~\cite{fang2021spirals}, coupled to some hard ferromagnet with a strong out-of-plane anisotropy.

\section*{Acknowledgment}
This work was supported by the grants of the Slovak Research and Development Agency (Grant No. APVV-20-0150) and the Scientific Grant Agency of Ministry of Education of Slovak Republic (Grant No. 1/0531/19).

\bibliographystyle{elsarticle-num}

\end{document}